\documentclass[12pt]{iopart}
\usepackage{iopams}
\usepackage{setstack}
\usepackage{graphicx} 
\usepackage{cite}

\begin{document}

\title{A skyrmion-based spin-torque nano-oscillator}

\author{F Garcia-Sanchez$^1$, J Sampaio$^{2,3}$, N Reyren$^2$, V Cros$^2$ and J-V Kim$^1$}
\address{$^1$Institut d'Electronique Fondamentale, CNRS, Univ. Paris-Sud, Universit\'e Paris-Saclay, 91405 Orsay, France}
\address{$^2$Unit{\'e} Mixte de Physique, CNRS, Thales, Univ. Paris-Sud, Universit{\'e} Paris-Saclay, 91767 Palaiseau, France}
\address{$^3$Laboratoire de Physique des Solides, CNRS, Univ. Paris-Sud, Universit{\'e} Paris-Saclay, 91405 Orsay, France}
\ead{joo-von.kim@u-psud.fr}

\begin{abstract}
A model for a spin-torque nano-oscillator based on the self-sustained oscillation of a magnetic skyrmion is presented. The system involves a circular nanopillar geometry comprising an ultrathin film free magnetic layer with a strong Dzyaloshinkii-Moriya interaction and a polariser layer with a vortex-like spin configuration. It is shown that spin-transfer torques due to current flow perpendicular to the film plane leads to skyrmion gyration that arises from a competition between geometric confinement due to boundary edges and the vortex-like polarisation of the spin torques. A phenomenology for such oscillations is developed and quantitative analysis using micromagnetics simulations is presented. It is also shown that weak disorder due to random anisotropy variations does not influence the main characteristics of the steady-state gyration.
\end{abstract}

\maketitle

\section{Introduction}
The flow of spin-polarised electrons through magnetic multilayers exerts torques on local magnetic moments and gives rise to a number of dynamical phenomena not accessible with magnetic fields alone. One example involves the self-sustained oscillation of the magnetisation, whereby the spin transfer torques in particular nanoscale geometries can compensate the intrinsic magnetic damping on average over a period of oscillation. Systems exhibiting such behaviour are referred to as spin-torque nano-oscillators (STNOs) and possible oscillation modes include spin wave eigenmodes in confined geometries such as pillars~\cite{Kiselev:2003hp}, self-localized spin wave bullet modes in extended films~\cite{Slavin:2005es}, magnetic vortices~\cite{Pribiag:2007dk, Pufall:2007jc, Mistral:2008js, Dussaux:2010ef}, and dynamical droplet solitons~\cite{Mohseni:2013eh}. STNOs have garnered much interest in recent years because their frequencies lie in the microwave range (0.1-100 GHz) and have been found to be tunable over a large range of applied magnetic fields and currents, which makes them attractive for applications as wide-band nanoscale electrical oscillators~\cite{Rippard:2004gy} and sensitive magnetic field sensors~\cite{Braganca:2010in}. From a more fundamental viewpoint, such systems are also interesting because they exhibit strongly nonlinear dynamics that can be used to study chaotic phenomena~\cite{PetitWatelot:2012be} and possible implementations of neuro-inspired computing~\cite{Nikonov:2013vw}.

A key feature of STNOs based on topological solitons, such as magnetic vortices, is the presence of a confining potential that governs the soliton dynamics. In magnetic nanopillars such as circular dots, the vortex state minimises the dipolar energy whereby the curling magnetisation state avoids the generation of magnetic charges at the dot edges and surfaces. As such, the dot geometry itself acts to confine the vortex toward the dot centre, where the effective potential for the vortex dynamics can be approximated as a quadratic function for small displacements about the centre~\cite{Ivanov:2007kn, Dussaux:2012cr}. In magnetic nanocontacts, on the other hand, the vortex resides in an extended multilayer film but is confined by the Zeeman energy associated with the Oersted-Ampere field generated by the electric current, which flows through a metallic contact 20-200 nm in diameter.  For this system the effective potential is better described by a linear function for vortex displacements relative to the nanocontact centre~\cite{Mistral:2008js}. In both cases, the nature of the potential is important because it not only determines the frequency of the self-sustained vortex gyration but also whether a current threshold exists for such oscillations~\cite{Mistral:2008js, Ivanov:2007kn, Dussaux:2012cr}.

Over the past few years, there has been increasing focus on another kind of topological soliton -- the magnetic skyrmion -- which appears in systems with a sizeable Dzyaloshinskii-Moriya interaction. This interaction, originally proposed to explain weak ferromagnetism in canted spin systems~\cite{Dzyaloshinsky:1958vq, Moriya:1960kc}, not only arises in non-centrosymmetric crystals such as the B20 class of materials MnSi and FeGe~\cite{Bak:1980fy}, but also in multilayered films lacking inversion symmetry~\cite{Bogdanov:2001hr} such as ultrathin ferromagnetic films in contact with a strong spin orbit material~\cite{Fert:1990, Yang:2015} like W(110)/Mn~\cite{Bode:2007em},  W(110)/Fe~\cite{Heide:2008da},  Ir (111)/Fe~\cite{Heinze:2011ic}, Pt/NiFe~\cite{Stashkevich:2015ku, Nembach:2015}, and Pt/Co~\cite{Thiaville:2012ia, Freimuth:2014en, Hrabec:2014hm, Di:2015fe, Tetienne:2015ka, Benitez:2015ba}. Much of the interest has been motivated by prospects of using skyrmions as nanoscale particles for information storage~\cite{Kiselev:2011cm, Fert:2013fq, Romming:2013iq}, where initial studies have suggested that these objects are less prone to defect-induced pinning~\cite{Fert:2013fq, Sampaio:2013kn, Iwasaki:2013hb} and can be moved by spin torques at lower current densities~\cite{Jonietz:2010fy}, which represent important advantages over similar schemes based on magnetic domain walls. Recent experiments have shown that skyrmions can be stabilized at room temperature in ultrathin film dots~\cite{MoreauLuchaire:2016, Boulle:2016}.

Like magnetic vortices, the dynamics of skyrmions is inherently gyrotropic, which suggests that they might be useful for applications as spin-torque nano-oscillators. In this article, we explore using micromagnetics simulations and analytical modelling the conditions under which self-sustained oscillations of skyrmions driven by spin transfer torques are possible in confined geometries such as ultrathin film circular dots. We show that the repulsion at boundary edges~\cite{Rohart:2013ef}, which underpins the resilience of skyrmions to edge defects, can act as a suitable confining potential for skyrmion oscillation if the orientation of the spin transfer torques possesses a vortex-like structure.

\section{Phenomenological model}
\begin{figure}
\centering\includegraphics[width=12cm]{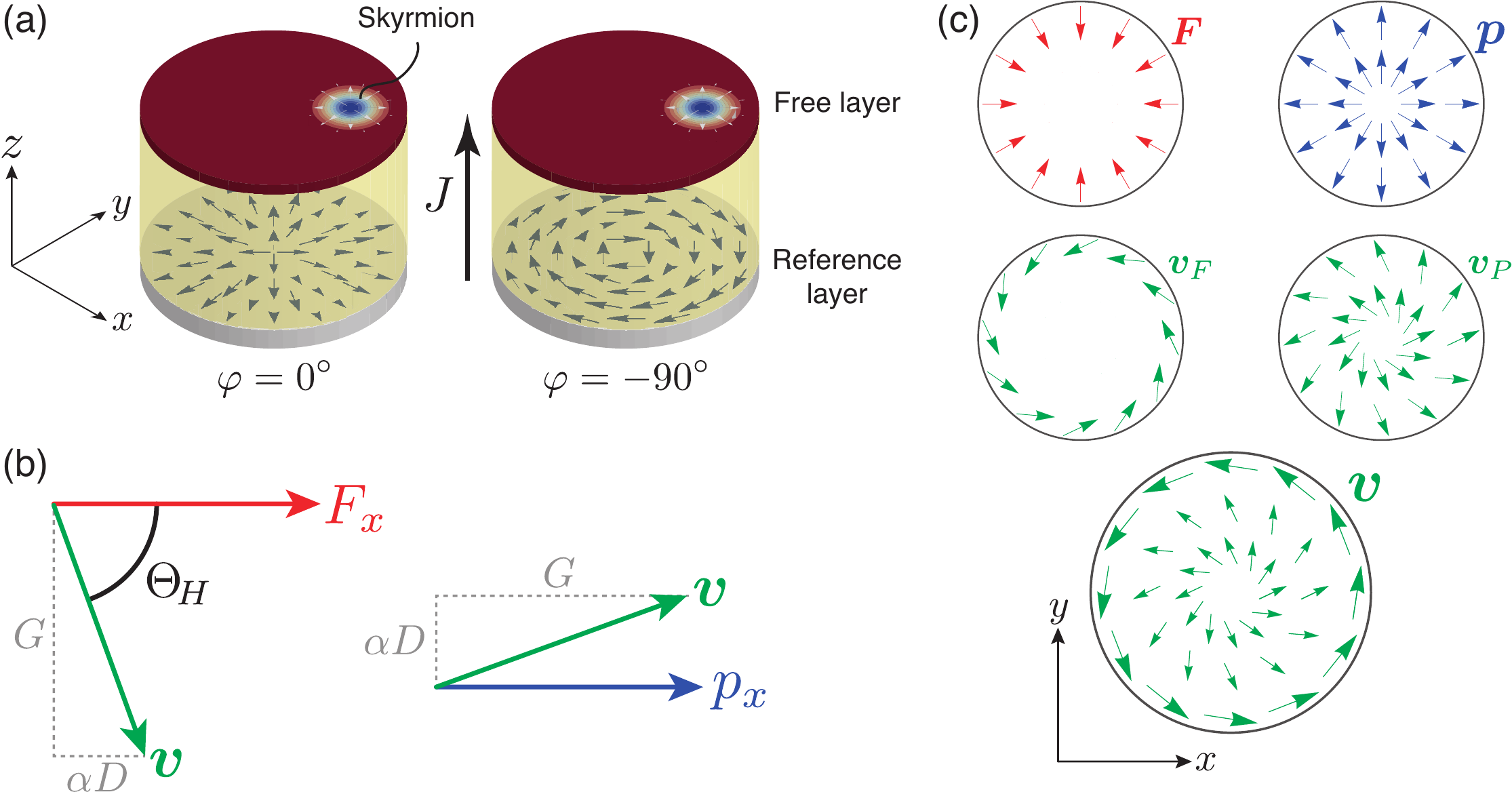}
\caption{Geometry and operating principle of a skyrmion-based spin-torque nano-oscillator. (a) Circular nanopillar with a gyrating skyrmion in the free layer and two possible vortex-like magnetic configurations $\varphi$ in the reference layer. The current $J$ flows perpendicular to the film plane along the $z$ direction. (b) Skyrmion velocity $\vec{v}$ in response to a uniform force $F_x$ and spin torques due to a uniform spin polarisation $P_x$, where $\Theta_H$ represents a Hall deflection angle, $G$ is the magnitude of the gyrovector, and $\alpha D$ represents the effective damping constant. (c) Phenomenological model illustrating how the potential force due to the boundary edge confinement ($\bi{F}$) and spin torques ($\bi{p}$) result in the velocity fields $\bi{v}_F$ and $\bi{v}_p$, respectively. The overall motion of the skyrmion is given by the resultant velocity field $\bi{v}$, where a limit cycle around the dot is expected.}
\label{fig:geometry}
\end{figure}

We consider a spin-torque nano-oscillator based on a circular nanopillar geometry as depicted in Fig.~\ref{fig:geometry}a. 
%
%
The film plane is defined by the $xy$-plane, while $z$ represents the direction perpendicular to the plane. The magnetic free layer is taken to be an ultrathin ferromagnet with perpendicular magnetic anisotropy and a Dzyaloshinskii-Moriya interaction such that a single isolated skyrmion represents a metastable energy state. Spin torques acting on this free layer are assumed to arise from a spin-polarised current flowing perpendicular to the film plane (CPP), where the spin polarisation takes the form of a vortex-like state in the in-plane magnetised reference layer. Vortices with a circulating magnetisation [e.g., $\varphi=-90^\circ$ in Fig.~\ref{fig:geometry}(a)] represent a micromagnetic ground state in ferromagnetic dots with certain aspect ratios~\cite{Guslienko:2008ty}, while those with a radial magnetisation configuration (e.g., $\varphi=0^\circ$) can appear in antiferromagnetically-coupled dots~\cite{Phatak:2012hn, Wintz:2013kf}. We also consider the influence of a static magnetic field applied along the perpendicular direction $z$.

The low-energy magnetisation dynamics in strong ferromagnets is well-described by the Landau-Lifshitz-Gilbert-Slonczewski (LLGS) equation,
\begin{equation}
\frac{d \bi{m}}{dt} = -\gamma \mu_0 \, \bi{m} \times \bi{H}_{\rm eff} + \alpha \bi{m} \times \frac{d \bi{m}}{dt} + \sigma J \bi{m} \times \left( \bi{m} \times \bi{p} \right),
\label{eq:LLGS}
\end{equation}
where $\bi{m} = \vec{M}/M_s$ is a normalised vector, $\| \bi{m} \| = 1$, and $M_s$ is the saturation magnetisation. The first term on the right hand describes magnetisation precession about the local effective field $\bi{H}_{\rm eff} = -(1/\mu_0 M_s) \delta U / \delta \bi{m}$, which represents a variational derivative of the free energy $U$ with respect to $\bi{m}$. The energy terms considered here are the isotropic exchange interaction, uniaxial anisotropy, dipole-dipole interactions, the interfacial form of the Dzyaloshinskii-Moriya interaction, and the Zeeman interaction associated with a static external magnetic field. The second term is the Gilbert term representing viscous damping, where $\alpha$ is the Gilbert damping constant. The third term is the Slonczewski term describing spin transfer torques, where $\sigma$ is an efficiency factor and $\bi{p} \equiv \bi{p}(\bi{r})$ represents the spin polarisation vector, which we consider to be spatially inhomogeneous. $J$ represents the current density, where in the convention used here (Fig.~\ref{fig:geometry}a) $J>0$ signifies the flow of conventional current from the reference layer to the free layer (i.e., electron flow from the free layer to the reference layer).

In order to outline the underlying physics involved in the skyrmion oscillations it is instructive to first consider the skyrmion dynamics within the Thiele approach, which assumes a rigid profile for the skyrmion core magnetisation and allows its dynamics to be described by its position in the film plane, $\bi{X}_0(t) = \left[X_0(t),Y_0(t)\right]$. The equations of motion are found by integrating out the additional degrees of freedom in (\ref{eq:LLGS}) to obtain
\begin{equation}
\bi{G} \times \dot{\bi{X}}_0 + \alpha D \dot{\bi{X}}_0 + \bi{\Gamma}_{\rm STT}(\bi{X}_0) = -\frac{\partial U}{\partial \bi{X}_0},
\label{eq:Thiele}
\end{equation}
where $\dot{\bi{X}}_0 \equiv d \bi{X}_0 / dt$. Let $\bi{m} = \left( \cos\phi \sin\theta, \sin\phi \sin\theta, \cos\theta \right)$, where $(\theta,\phi)$ are the polar and azimuthal angles, respectively, of the magnetisation unit vector in spherical coordinates. The first term on the left-hand side of (\ref{eq:Thiele}) is the gyrotropic term, where the gyrovector is defined by 
\begin{equation}
\bi{G} = \frac{M_s}{\gamma} \int dV \, \sin(\theta) \left( \nabla \theta \times \nabla \phi \right),
\end{equation}
and the second term is the Gilbert damping, where $D\equiv D_{ii}$ is the diagonal component of the damping tensor~\cite{Thiele:1973br}, 
\begin{equation}
D_{ij} = \frac{M_s}{\gamma} \int dV \, \left[ \frac{\partial \theta}{\partial x_i}  \frac{\partial \theta}{\partial x_j} + \sin^2(\theta) \, \frac{\partial \phi}{\partial x_i}  \frac{\partial \phi}{\partial x_j} \right],
\end{equation}
with $x_i$ representing the cartesian coordinates $(x,y,z)$. The third term on the left-hand side represents the spin torque term, which is a dissipative torque similar to the Gilbert damping. This torque term is given by $\bi{\Gamma}_{\rm STT} = \partial W_{\rm STT} / \partial \dot{\bi{X}}_0$, which is derived from the Rayleigh dissipation function~\cite{Sampaio:2013kn, Khvalkovskiy:2009be, Consolo:2011go, Kim:2012du}, 
\begin{equation}
W_{\rm STT} = -\sigma I \int dV \, \bi{p} \cdot \left( \bi{m} \times \frac{d \bi{m}}{dt} \right).
\end{equation}
Finally, the term on the right hand side represents a conservative force acting on the skyrmion core that is derived from the total free energy of the system. It is interesting to note that (\ref{eq:Thiele}) describes a dynamics that is inherently non-Newtonian, since the response of the skyrmion core to a conservative force is gyrotropic and in a direction perpendicular to the force, rather than a linear acceleration in a direction collinear with the force. This gyrotropic dynamics underpins the skyrmion oscillations that we consider here.

To better understand the dynamics described by (\ref{eq:Thiele}), it is useful to examine the response of the skyrmion core to uniform forces and spin-torques, as illustrated in Fig.~\ref{fig:geometry}b. Let us assume that the skyrmion core magnetisation points along the $+z$ direction, while the uniform background magnetisation is oriented along the $-z$ direction; this leads to $\bi{G} = \left( 4\pi M_s d / \gamma \right)\hat{\bi{z}}$ in the absence of an applied magnetic field $B_z = 0$, where $d$ is the film thickness. For a conservative force along the $x$ direction, $F_x$, it follows from (\ref{eq:Thiele}) that the resultant motion $\bi{v}$ is nearly perpendicular to this force, where $\Theta_H = \tan^{-1}\left[G / (\alpha D) \right]$ represents a deflection angle that is analogous to a Hall effect angle. One notes that the presence of the Gilbert damping results in a finite Newtonian response to the applied force. For spin torques associated with a uniform spin polarisation along $x$, $P_x$, and $J >0$, the resultant motion is largely along the $x$ direction with a deflection along $y$ at an angle $90^{\circ} - \Theta_H$. From the basis of these results, we can anticipate how the current-driven dynamics in a circular dot in Fig.~\ref{fig:geometry}a might look like for the radial vortex polariser, $\varphi = 0^{\circ}$. First, the dot edges provide a repulsive potential that confines the skyrmion, which leads to a radial inward force that only exists near the boundary (Fig.~\ref{fig:geometry}c). A circular motion results from this force and is illustrated by the velocity field $\bi{v}_F$. Second, the radial polariser drives a motion outwards from the dot circle with a small azimuthal component, as illustrated by the velocity field $\bi{v}_P$. In contrast to the boundary force, the spin-torques act everywhere inside the dot with the same magnitude because it is assumed that the current density is uniform across the lateral surface of the dot. By combining the velocity fields from the boundary force and spin-torques, the resultant motion is a steady state gyration of the skyrmion core near the edge of the dot in a counterclockwise rotation with the configuration considered here ($G > 0$), as shown by the field $\bi{v}$ in (Fig.~\ref{fig:geometry}c). As such, we can expect similar dynamics to oscillations described by a limit cycle, where paths from any initial condition converge to a stable orbit. It is important to note that such steady-state oscillations are only possible for one current polarity, since the velocity field for the opposite current polarity is reversed (but the boundary forces remain unaffected) and results in a stable fixed point for the skyrmion core at the dot centre. Moreover, no threshold current for self-oscillations is predicted in this picture.

\section{Micromagnetics Simulations}

In order to test whether this phenomenological model is accurate, we solved numerically the full dynamical system given in (\ref{eq:LLGS}) using the micromagnetics simulation codes \texttt{MuMax2} and \texttt{MuMax3}~\cite{Vansteenkiste:2014et}. The codes perform a time integration of the coupled equations of motion using a finite-difference scheme to discretise the magnetisation vector $\bi{m}(\bi{r})$. For the majority of the results presented here, we considered a circular dot with a diameter of 150 nm and a thickness of 0.6 nm, which was discretised using $128 \times 128 \times 1$ finite difference cells. The materials parameters were chosen to represent the 0.6 nm thick Pt/Co/AlOx system, which possesses a large Dzyaloshinskii-Moriya constant~\cite{Pizzini:2014bf, Belmeguenai:2015hj, Cho:2015eq}; we assumed an exchange constant of $A = 16$ pJ/m, a uniaxial anisotropy of $K_u = 1.26$ MJ/m$^3$, a saturation magnetisation of $M_s = 1.1$ MA/m, a Dzyaloshinskii-Moriya constant of $D_{\rm ex} = 2.5$ mJ/m$^2$, and a Gilbert damping constant of $\alpha = 0.3$~\cite{Schellekens:2013it}. For the spin transfer torques, we assume a spin polarization of $P = 0.5$ and a symmetric form for the Slonczewski term ($\lambda = 1.0$ ~\cite{Vansteenkiste:2014et}).

\begin{figure}
\centering\includegraphics[width=10cm]{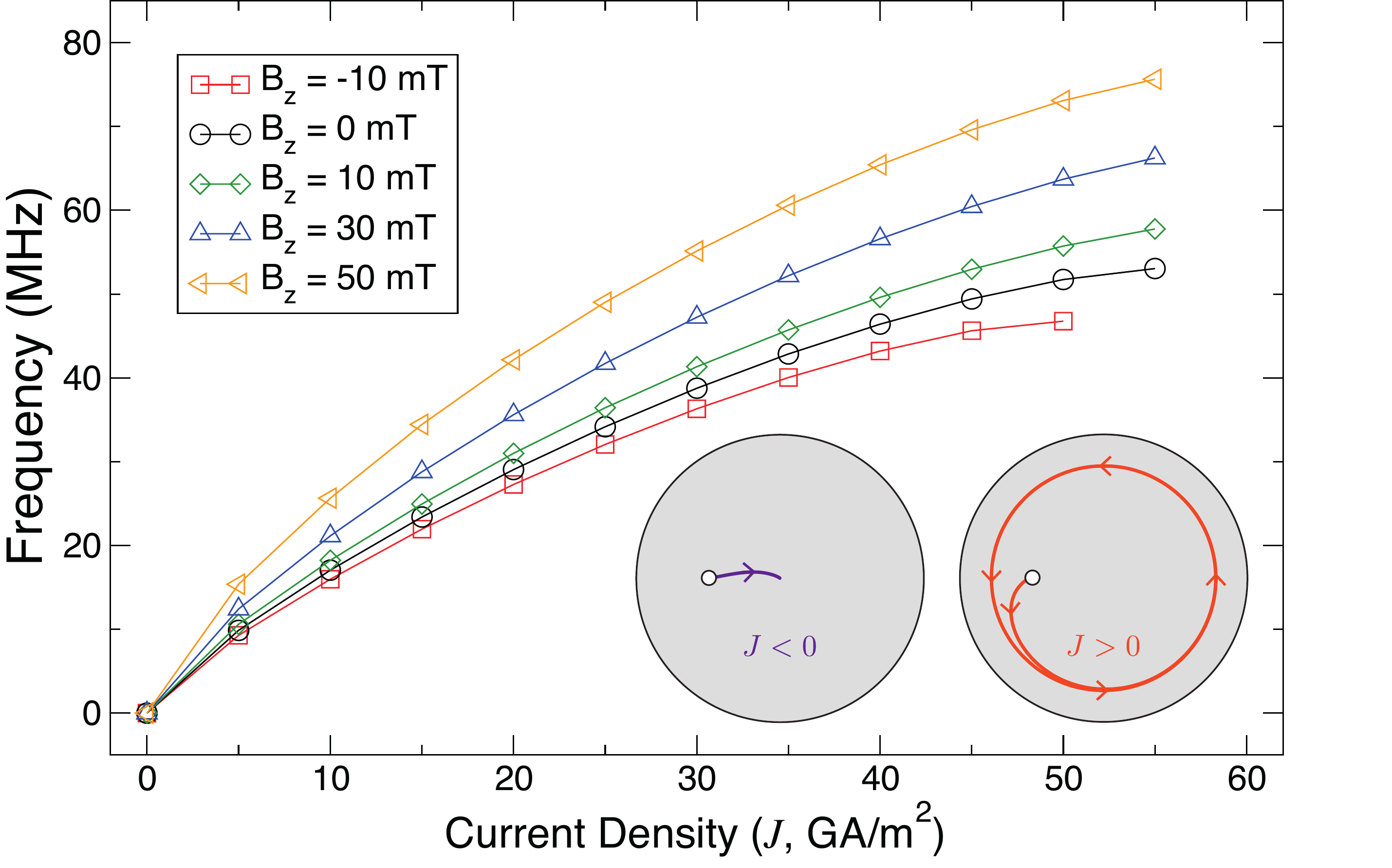}
\caption{Current dependence of the skyrmion gyration frequency for different perpendicular applied fields, $B_z$. An upper critical current density exists at which the skyrmion is expelled from the dot. The inset shows the transient trajectories of the skyrmion core for opposite current polarities.}
\label{fig:fvsI}
\end{figure}

In Fig.~\ref{fig:fvsI}, we show the current dependence of the skyrmion gyration frequency for different values of the perpendicular applied field, $B_z$.
%
%
This field changes the size of the skyrmion core~\cite{Kiselev:2011cm, Sampaio:2013kn, Kim:2014cm, Romming:2015il}. Here, we considered a polariser with $\bi{p}(\bi{r})$ pointing radially outward from the dot centre, as illustrated in Fig.~\ref{fig:geometry}(a) for $\varphi = 0^\circ$. In the inset of the figure, the transient dynamics of the skyrmion core are shown, with an initial position off-centred. For negative current densities, the spin torques drive the skyrmion core with a clockwise gyration toward the dot centre, where it remains immobile indefinitely. This follows from the picture given in Fig.~\ref{fig:geometry}(c), where the resultant spin torque polarisation  $\bi{p}$ is oriented radially inwards toward the dot centre for negative currents, resulting in an inward spiralling motion of the skyrmion core in a clockwise direction.  On the other hand, positive currents drive the core outwards from the dot centre toward the boundaries and reaches a steady state gyration in a counterclockwise sense, as expected from the phenomenology depicted in Fig~\ref{fig:geometry}. Note there is no critical current for the onset of oscillations, which follows from the Thiele approach described earlier. However, an upper critical current is found between 50 and 60 GA/m$^2$ at which the skyrmion is expelled from the dot. This occurs because the confining potential due to the repulsion from the partial domain walls at the boundary~\cite{Rohart:2013ef} is no longer sufficiently strong to counteract the radial component of the spin-torque induced force on the skyrmion core that propels it outwards.

\begin{figure}
\centering\includegraphics[width=10cm]{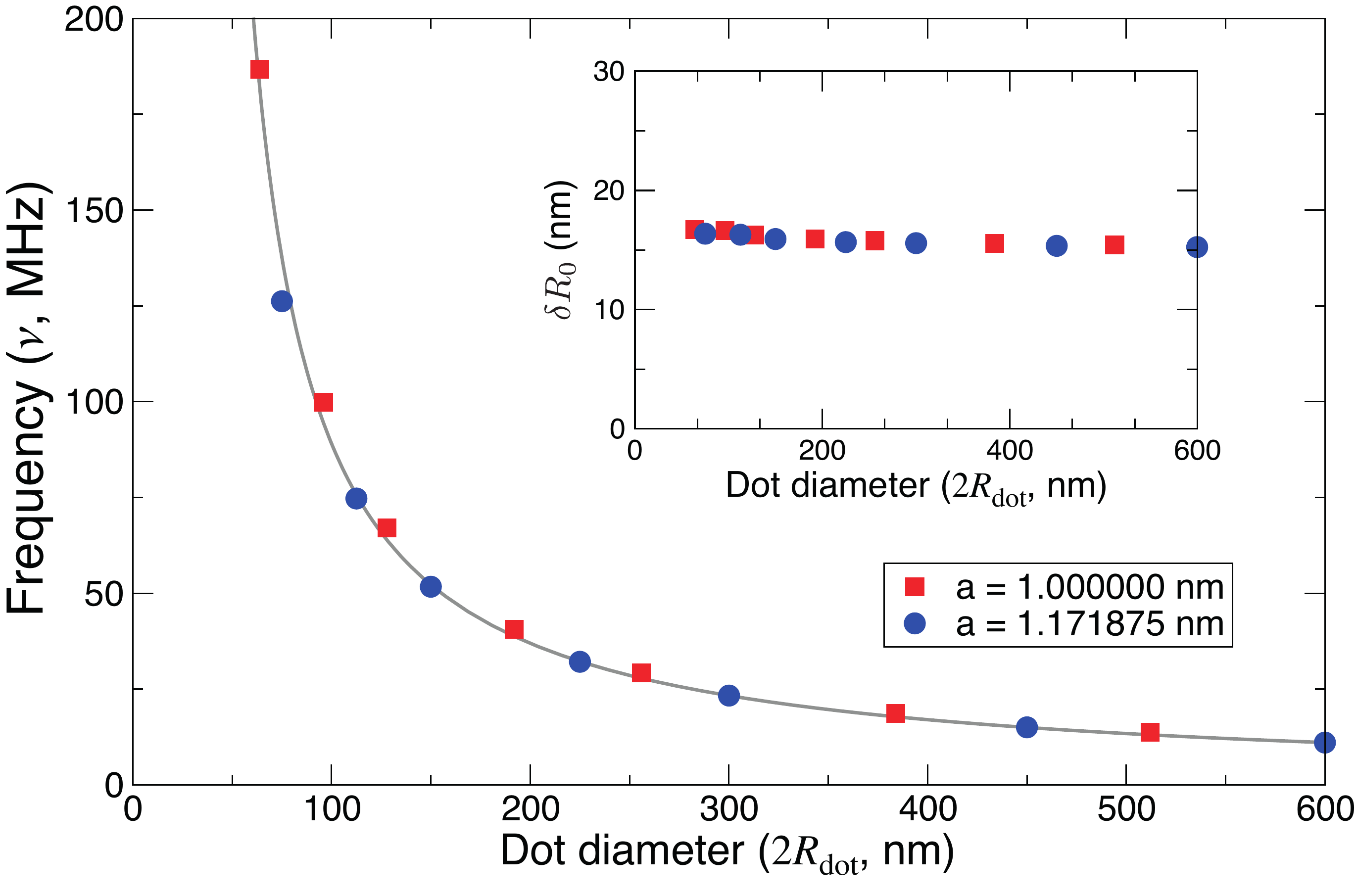}
\caption{Dependence of the skyrmion gyration frequency on dot diameter for $J = 50$ GA/m$^2$ and $B_z = 0$ mT, with two different finite difference cell sizes $a$. Note that $1.171875 = 150/128$, which is the cell size used for most of the results presented here. The inset shows the radial distance of the skyrmion gyration orbit from the dot edge, $\delta R_0$, as a function of dot diameter. The solid line represents a fit of the gyration frequency, $\nu$, using the function, $\nu = C/(R_{\rm dot} - \delta R_0 )$ where $C$ is a constant and $R_{\rm dot}$ is the dot radius.}
\label{fig:fvsR}
\end{figure}

Based on the arguments put forward in the previous section, the radius of the steady state orbit should only be determined by the dot radius, since the restoring force associated with the confining potential only acts at the boundary edge of the dot. In order to test this hypothesis, we examined the dependence of the gyration frequency on the dot size for the same polariser configuration in Fig.~\ref{fig:fvsI}. As the results in Fig.~\ref{fig:fvsR} show, the gyration frequency is found to be inversely proportional to the dot diameter, where the steady state gyration orbit, $R_0 \equiv \| \bi{X}_0 \|$ (taking the origin at the center of the dot), is found to scale with the dot size. 
%
%
This can be seen in the inset of Fig.~\ref{fig:fvsR}, where the radial distance between the gyration orbit and the dot edge, $\delta R_0$, is found to be largely independent of the dot size. We also performed simulations for a slightly different finite cell size (1 nm) in order to obtain a larger set of data points, where it is found that the additional data fall on the same inverse dependence of the gyration frequency on the dot diameter. We note, however, that the oscillatory dynamics described here are not seen in smaller dot sizes where the core radius becomes comparable to the dot dimensions ($R_0 \lesssim 25$ nm). At these sizes, the spatial variation of the spin polarisation plays a greater role, where it is observed that a finite threshold current is required for self oscillation. We will not give any further consideration to this regime here but instead focus on larger systems in which the skyrmion oscillation is well established.

\begin{figure}
\centering\includegraphics[width=14cm]{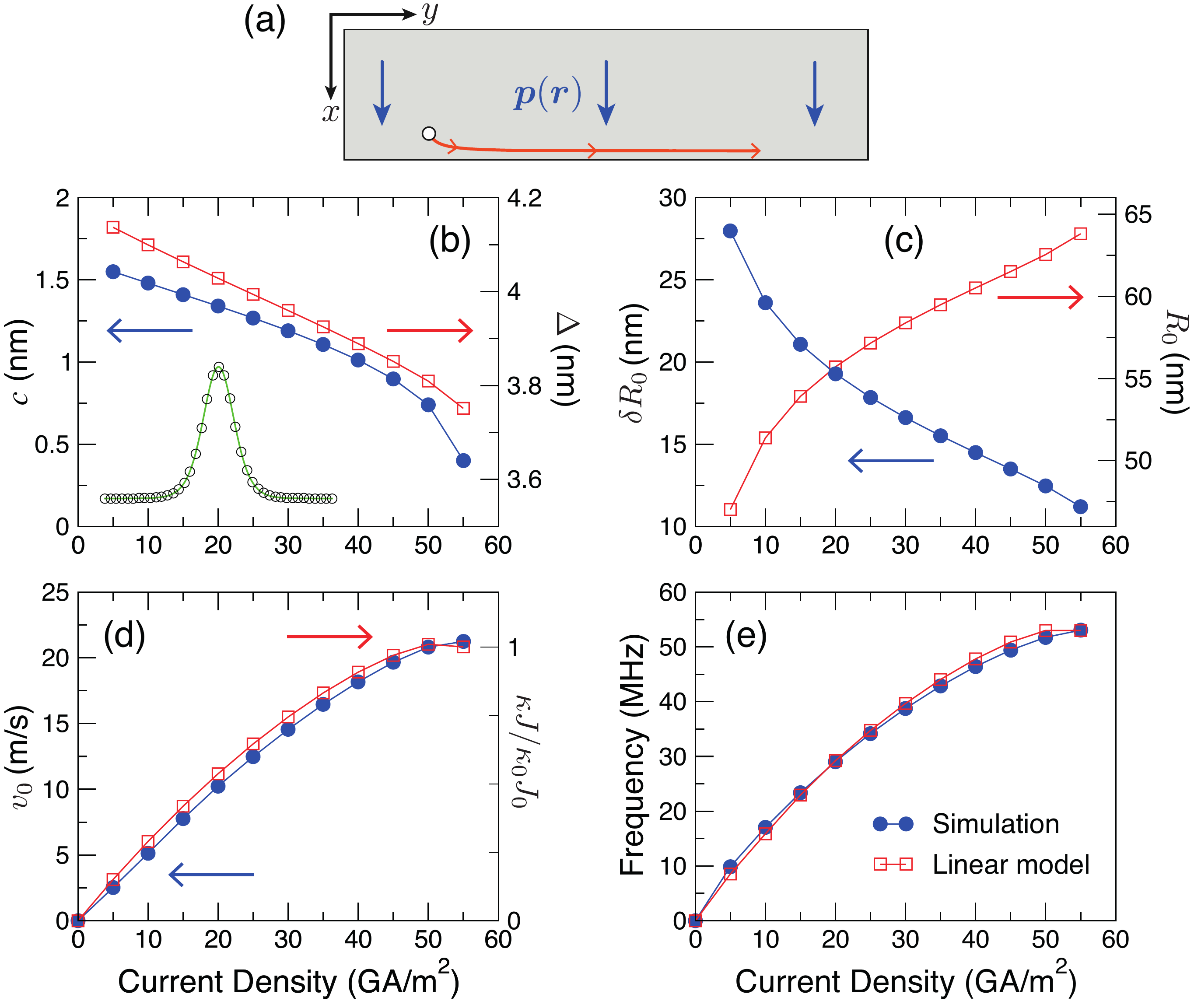}
\caption{Characteristics of linear skyrmion propagation along a straight boundary edge. (a) Schematic illustration of skyrmion motion along the $y$ direction as a result of Slonczewski spin-torques with $\bi{p} = \hat{\bi{x}}$. (b) Current dependence of the fitted parameters $c$ and $\Delta$ of the double-wall \emph{ansatz} for the skyrmion core profile. The inset shows the fitted profile with the points determined from simulation. (c) Current dependence of the separation distance between the core and the boundary edge, $\delta R_0$, from which the expected gyration radius for a 150 nm diameter dot, $R_0$, is derived. (d) Current dependence of the linear skyrmion velocity, $v_0$, and the spin-torque factor, $\kappa J$, normalised to its value at $J=55$ GA/m$^2$, $\kappa_0 J_0$. (e) Comparison between the simulated and predicted (model) current dependence of the gyration frequency for $B_z = 0$ mT (\emph{cf}. Fig.~\ref{fig:fvsI}).}
\label{fig:params}
\end{figure}

Based on the phenomenology described in Section 2 and the results in Fig.~\ref{fig:fvsR}, one might expect a simple linear dependence of the gyration frequency as a function of applied current density. However, as the results in Fig.~\ref{fig:fvsI} highlight, the current dependence observed in simulation is nonlinear. To elucidate the origin of this nonlinearity, we revisited a simpler system by examining the dynamics of an isolated skyrmion in a rectangular wire with a uniform spin polarisation vector $\bi{p}$. To mimic the gyration in Fig.~\ref{fig:fvsI}, we assumed $\bi{p}(\bi{r}) = \hat{\bi{x}}$ and examined the linear propagation parallel to the boundary edge along the $y$ axis, as shown in Fig.~\ref{fig:params}(a), as a function of applied current density.
%
%
This represents the limiting case of an infinitely large circular dot where the local curvature of the dot edge can be neglected and the spin polarisation is uniform on the length scale of the skyrmion core. We find a number of features that contribute to the nonlinearity of the current dependence of the gyration frequency. First, the skyrmion core is found to shrink as the current density is increased. This is shown in Fig.~\ref{fig:params}(b), where the parameters $c$ and $\Delta$, obtained by fitting the perpendicular magnetisation component $m_z$ using the function
\begin{equation}
m_z(r) =  \frac{4 \cosh^2{c}}{\cosh{2c} + \cosh{\left( 2r/\Delta \right)}} - 1,
\label{eq:core}
\end{equation}
are presented as a function of applied current density. This function is based on a double-soliton \emph{ansatz} that describes two successive 180$^\circ$ homochiral domain walls with a characteristic length $\Delta$ that are separated by a distance $c$~\cite{Braun:1994ff}. It has been shown elsewhere that this function provides a good quantitative description of the skyrmion core profile measured in low-temperature experiments~\cite{Romming:2015il}. An example of a fit of to the simulation core profile using (\ref{eq:core}) is given in the inset of Fig.~\ref{fig:params}(b). As the current is increased, we observe a decrease in both the characteristic length $\Delta$ and the separation distance $c$. Second, the combination of these effects leads to a decrease in the separation distance between the core and the boundary edge, $\delta R_0$, as shown in Fig.~\ref{fig:params}(c). From this quantity, we can obtain an accurate estimate of the radius of gyration for any dot size based on the results in the inset of Fig.~\ref{fig:fvsR}, where $\delta R_0$ was found to be independent of the dot size; the expected variation for a dot diameter of 150 nm is presented in Fig. \ref{fig:params}(c). Third, the linear velocity of the skyrmion along the boundary is also found to exhibit a similar nonlinear variation as a function of the applied current. This behaviour can be understood in terms of the reduction in the skyrmion core size, which leads to a decrease in the spin torque efficiency. To understand this quantitatively, we can evaluate the dissipation function $W_{\rm STT}$ associated with the spin torques by assuming a rigid core approximation, $\partial_t \bi{m}(\bi{r},t) = -\dot{\bi{X}}_0 \cdot \bi{\nabla} \bi{m}(\bi{r},t)$, which for $\bi{p} = \hat{\bi{x}}$ leads to an integral of the form,
\begin{equation}
W_{\rm STT} = \pi \sigma J \: \dot{Y}_0 \int_{0}^{\infty} dr \, \left( r \frac{\partial \theta}{
\partial r} + \sin{\theta} \cos{\theta}   \right) \equiv \sigma J \: \dot{Y}_0 \: \kappa,
\label{eq:dissfnpolar}
\end{equation}
where $\theta(r)$ represents the polar angle of the magnetisation vector $\bi{m}$ in spherical coordinates and $r$ is the radial spatial coordinate whose origin is the skyrmion core centre. As the core shrinks with increasing current, the integral term $\kappa$ in (\ref{eq:dissfnpolar}) also acquires a current dependence by virtue of the changes in $c$ and $\Delta$, as shown in Fig.~\ref{fig:params}(b). As such, the overall efficiency of the spin torques is no longer independent of the applied current. The spin torque term, $\kappa J$, normalised by its value at $J = 55$ GA/m$^2$, $\kappa_0 J_0$, is shown as a function of applied current in Fig.~\ref{fig:params}(d). Instead of the linear variation expected for a constant $\kappa$, the spin torque term exhibits a variation that closely mirrors the velocity versus current curve. This highlights the importance of the changes in the core profile on the current-driven edge dynamics of the skyrmion. Finally, we can estimate the expected current dependence of the gyration frequency in a circular dot with $\nu = v_0 / (2 \pi R_0)$ and by using the current dependence of $R_0$ and $v_0$ computed for a linear propagation along a straight boundary edge as shown in Figs.~\ref{fig:params}(b) and \ref{fig:params}(d), respectively. The result is shown in Fig.~\ref{fig:params}(e) (``linear model''), which is compared with the gyration frequency computed previously in Fig.~\ref{fig:fvsI} for $B_z = 0$ mT (``simulation''). An excellent quantitative agreement is found between the full simulation results and the estimates obtained by examining the linear propagation.

\begin{figure}
\centering\includegraphics[width=10cm]{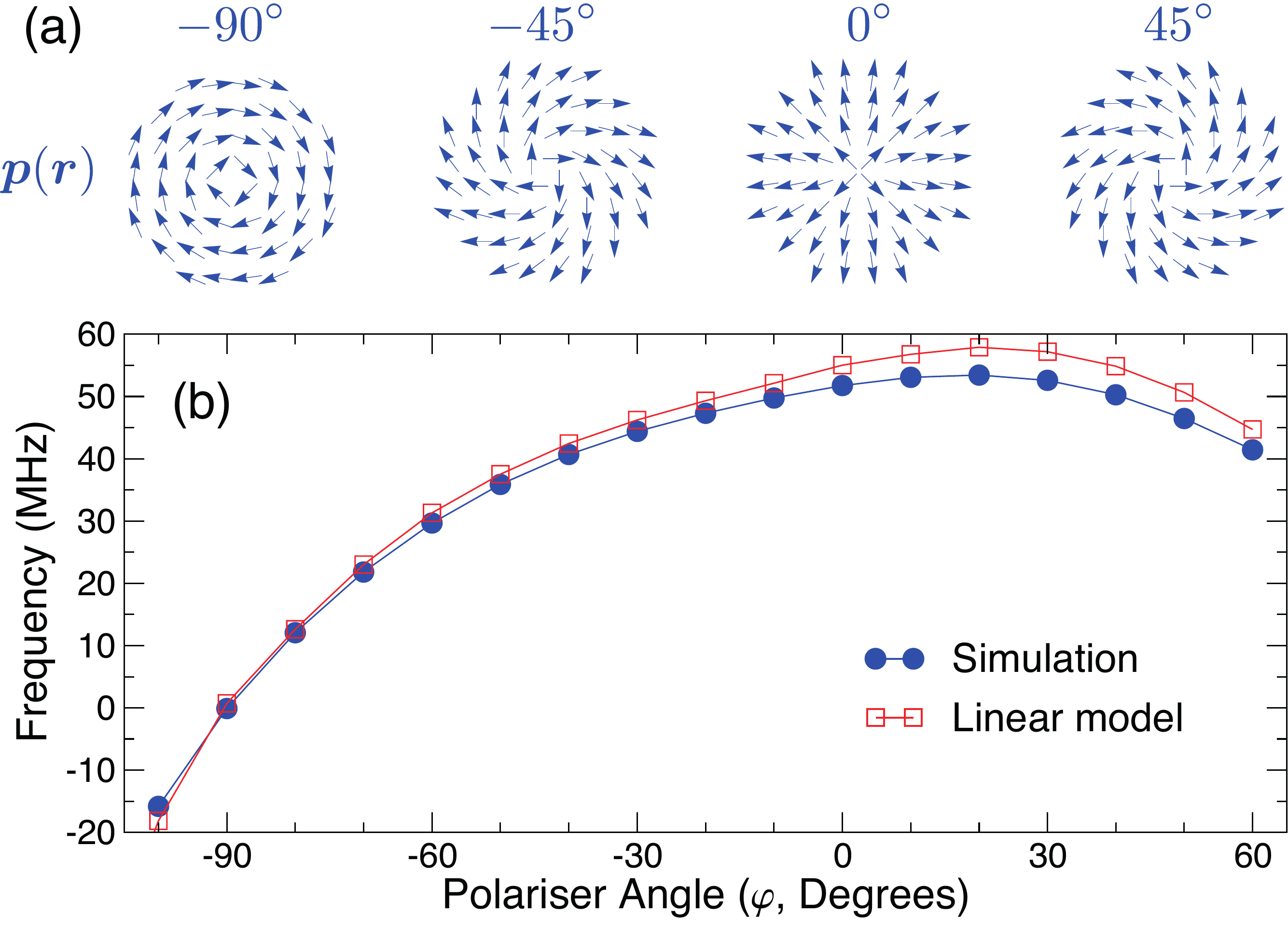}
\caption{Gyration frequency as a function of polariser angle. (a) Example of four different vortex-like polariser configurations, $\bi{p}(\bi{r})$. (b) Dependence of the gyration frequency on the polariser angle for an applied current density of $J = 50$ GA/m$^2$ and zero applied field, $B_z = 0$ mT. The blue dots represent data computed from simulations of the skyrmion core gyration in a circular dot (``simulation''), while the red squares represent the predicted frequency from the linear model (\emph{cf}. Fig.~\ref{fig:params}).}
\label{fig:fvspolangle}
\end{figure}

We have also examined the effect of different vortex-like configurations for the polarisation vector, which can be characterised by the angle $\varphi$ as shown in Fig.~\ref{fig:fvspolangle}(a). This angle determines the azimuthal component of the polariser through the relation $\bi{p} = (\cos{\phi_p}, \sin{\phi_p}, 0)$, where $\phi_p = \tan^{-1}\left( y/x \right) + \varphi$ and $(x,y)$ are spatial coordinates in the film plane with the origin at the dot centre. The dependence of the gyration frequency on the polariser angle $\varphi$ is shown in Fig.~\ref{fig:fvspolangle}(b) for a fixed current density of $J = 50$ GA/m$^2$ and under zero applied magnetic field.
%
%
The range of polariser angles studied was chosen so that steady state gyration is favoured, i.e., such that the spin torques drive skyrmion motion radially outward from the dot centre. We observe a large variation in the gyration frequency, with a maximum attained at around $\varphi = 20^{\circ}$ and a change in sign at around $\varphi = -90^{\circ}$. For the latter, the spin torques and the restoring force from the boundary edge drive the azimuthal motion of the skyrmion gyration in different directions but are equal in magnitude, which results in the skyrmion being driven to the dot boundary but with a vanishing frequency of gyration. The overall $\varphi$-dependence of the gyration frequency can be understood in terms of the linear model presented in Fig.~\ref{fig:params}, where the analysis is performed for different orientations of $\bi{p}$ with respect to the boundary edge rather than the current density. As Fig.~\ref{fig:fvspolangle} shows, the predicted frequency variation from the linear model accounts well for the simulated gyration frequency.

%
\section{Role of weak disorder}
The results of micromagnetics simulations presented up until this point have been conducted for a system free of material defects and edge roughness. The presence of such defects is important because spatial variations in the material parameters could induce fluctuations in the gyration orbit, which in turn would translate into an athermal component of the power spectrum of the skyrmion oscillations. The effect of edge roughness has been explored in previous work~\cite{Fert:2013fq, Sampaio:2013kn}, where it was demonstrated that edge repulsion effects, which govern skyrmion confinement~\cite{Rohart:2013ef}, can be sufficiently large to guided the skyrmion core around defects, much in the same way that confinement acts to maintain steady-state gyration in the dots considered here. 

\begin{figure}
\centering\includegraphics[width=14cm]{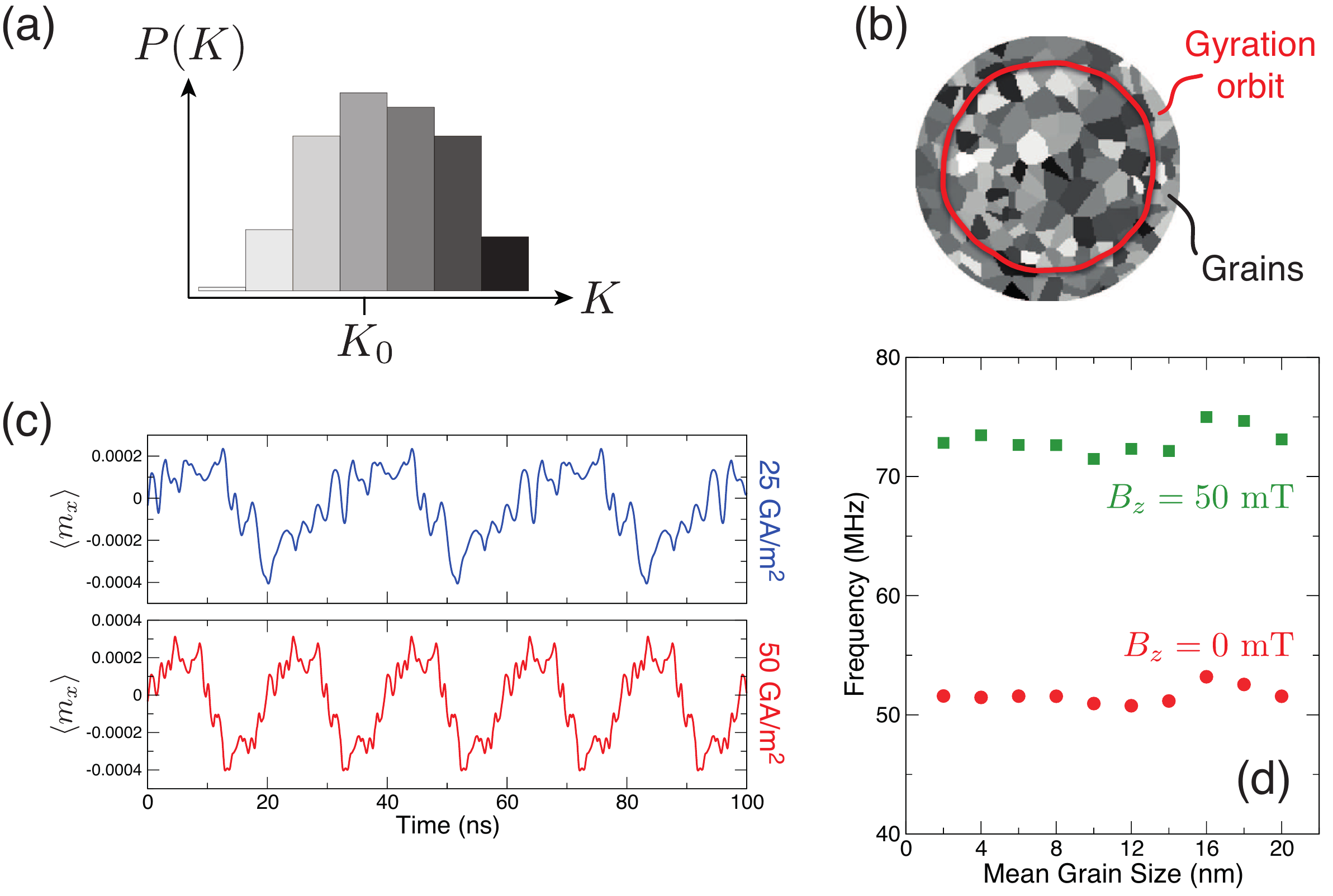}
\caption{Effect of a random anisotropy distribution on skyrmion gyration. (a) Distribution of anisotropy values $K$, which are drawn from a Gaussian distribution centred about $K_0$ with a standard deviation of $0.01 K_0$. (b) Grain distribution (grayscale) in the circular dot with an average grain size of 10 nm, where each grain is assigned a value of $K$ shown in (a). A steady-state gyration orbit is shown superimposed in red. (c) Mean value of the $x$ component of the magnetisation as a function of time, $\langle m_x(t) \rangle$, for two different applied currents $J$ at zero field and with the grain distribution shown in (b). Steady-state gyration is observed with no perceptible stochasticity. (d) Dependence of the gyration frequency on mean grain size at $J=50$ GA/m$^2$ under two different applied fields, $B_z$.}
\label{fig:disorder}
\end{figure}

Here, we have examined the role of a different kind of defect. We assume that the ultrathin ferromagnetic free layer is composed of an ensemble of grains, where each grain $i$ is assigned a local value of the uniaxial anisotropy constant, $K_{u,i}$, that is drawn from a Gaussian distribution centred at $K_{u,0} = 1.26$ MJ/m$^3$ with a standard deviation of $0.01 K_{u,0}$ [Fig.~\ref{fig:disorder}(a)]. 
%
%
An illustration of the grain structure for a 150 nm-diameter circular dot is given in Fig.~\ref{fig:disorder}(b), where the mean grain size is 10 nm and the grains are constructed using Voronoi tessellation. The grayscale in Fig.~\ref{fig:disorder}(b) corresponds to the anisotropy distribution in Fig.~\ref{fig:disorder}(a).

The presence of such an anisotropy distribution leads to local fluctuations in the effective magnetic potential seen by the skyrmion core, which enters as an additional component of $U$ on the right-hand side of (\ref{eq:Thiele}). The effect of such fluctuations can be seen in Fig~\ref{fig:disorder}(c), where the time evolution of the average in-plane magnetisation component, $\langle m_{x}(t)\rangle$, is presented for two values of the applied current for the grain structure shown in Fig.~\ref{fig:disorder}(b). We note the appearance of higher frequency fluctuations, due to the anisotropy distribution, on top of a background sinusoidal variation that represents the overall core gyration around the dot. Interestingly the time traces appear to be periodic without any perceptible noise, which can clearly be seen for the repeating motifs for $J = 50$ GA/m$^2$. The core trajectory corresponding to this applied current density is illustrated by the red curve in Fig.~\ref{fig:disorder}(b) that is superimposed on the grain structure. While deviations from the perfect circular trajectory can be seen, we note that successive orbits around the dot follow the same trajectory (within the spatial resolution of our simulations).

Fig.~\ref{fig:disorder}(d) shows how the gyration frequency varies with the mean grain size for two different applied magnetic fields at a constant current density of $J = 50$ GA/m$^2$. The gyration frequency is found to be largely independent of the mean grain size for both field values considered. We note that under $B_z = 50$ mT the skyrmion core radius increases to approximately 14 nm, compared with 10 nm at zero field, but this difference in size has no noticeable effect with respect to the grain structure considered. This suggests that results obtained for the defect-free case, along with the theoretical picture developed for it, should remain applicable for realistic systems in which a grain structure and weak disorder is present.

\section{Conclusion and outlook}
In conclusion, we have described a theoretical model for a spin-torque nano-oscillator based on magnetic skyrmions. The oscillator is based on the self-sustained gyration of the skyrmion that arises from a competition between geometric confinement due to boundary edges and an inhomogeneous spin polariser with a vortex-like configuration. In contrast to other known spin-torque nano-oscillators, such as vortex-based systems, there is no threshold current for the onset of oscillations for dots larger than the skyrmion core but there exists a critical current above which the skyrmion is expelled from the system. This leads to a finite range of applied current densities in which skyrmion gyration can take place (0 to 55-60 GA/m$^2$ with the parameters considered here), with gyration frequencies in the range of tens of MHz. The oscillatory dynamics has also been shown to be relatively robust to weak disorder in the form of random anisotropy fluctuations, but the effect of stronger disorder remains to be explored.

While the gyration frequencies described here are an order of magnitude lower than those of typical vortex-based spin-torque nano-oscillators, the skyrmion oscillator has the potential to offer richer dynamics. Since skyrmions are naturally compact and behave like point particles, the system could be extended to host multiple skyrmions in a straightforward manner, in the same spirit as the magnetic skyrmion racetrack but instead in an oscillator geometry. Depending on how the signal is read out, different skyrmion numbers and arrangements could allow complex periodic waveforms to be tailored. Such a device might not only be useful for signal generation, but also for neuro-inspired applications in which complex transient states can be exploited for information processing~\cite{Appeltant:2011jy}.

\section*{Acknowledgements}
The authors would like to acknowledge fruitful discussions with Constance Moreau-Luchaire and Albert Fert. This work was partially supported by the Agence Nationale de la Recherche (France), under grant agreement No. ANR-14-CE26-0012 (Ultrasky), and by the Horizon2020 Framework Programme of the European Commission, under grant agreement No. 665095 (MAGicSky).

\section*{References}
\bibliographystyle{iopart-num}
\bibliography{articles}

\end{document}